\documentclass[reprint]{revtex4-2}
\pdfoutput=1
\usepackage{graphicx}
\usepackage{dcolumn}
\usepackage{bm}
\usepackage{color}
\usepackage{amsmath}
\usepackage{amssymb}
\usepackage{comment}
\usepackage[utf8]{inputenc}
\usepackage{textcomp}
\usepackage{soul}
\usepackage{hyperref}
\usepackage{xr}
\usepackage[symbol]{footmisc}

\usepackage{subfiles}
\usepackage{blindtext}
\usepackage{amsmath,amsfonts,amsthm,bm} 
\usepackage{todonotes}
 
\usepackage{float}
\newfloat{suppfig}{tbh}{losf}
\floatname{suppfig}{Supplementary Figure}
\usepackage[a4paper, total={6.9in, 10.02in}]{geometry}
\usepackage{float}
\newcommand\emailx[1]{%
\move@AF%
\def\@affil{{\normalfont\,#1\strut}{}}%
}%

\makeatletter
\newcommand{\fmarki}{*}
\newcommand{\fmarkii}{\ensuremath{\dagger}}
\def\@fnsymbol#1{{\ifcase#1\or \fmarki\or \fmarkii\else\@ctrerr\fi}}
\makeatother
\renewcommand{\fmarki}{*}
\renewcommand{\fmarkii}{*}

\begin{document}

\title{Single-sample image-fusion upsampling of fluorescence lifetime images}


\author{V.Kapitány$^{\dagger,1,*}$}
\author{A.Fatima$^{\dagger,1}$}
\author{V.Zickus$^{1,2}$}
\author{J.Whitelaw$^{3,4}$}
\author{E.McGhee$^{1,3}$}
\author{R.Insall$^{3,5}$}
\author{L.Machesky$^{3,6}$}
\author{D.Faccio$^{1,*}$}

\affiliation{$^1$ School of Physics \& Astronomy, University of Glasgow, Glasgow G12 8QQ, UK}
\affiliation{$^2$ Department of Laser Technologies, Center for Physical Sciences and Technology, LT-10257, Vilnius, Lithuania}
\affiliation{$^3$Cancer Research UK, Beatson Institute, Glasgow, United Kingdom}
\affiliation{$^4$Present: School of Health \& Life Sciences, University of the West of Scotland, Blantyre G72 0LH}
\affiliation{$^5$Present: Department of Cell and Developmental Biology, University College London, London WC1E 6BT}
\affiliation{$^6$Present: Department of Biochemistry, University of Cambridge, Cambridge CB2 1GA}
\affiliation{$^{\dagger}$These authors contributed equally to the work}
\affiliation{$^{*}$ Corresponding authors: \href{mailto:valentin.kapitany@glasgow.ac.uk}{valentin.kapitany@glasgow.ac.uk}, \href{mailto:daniele.faccio@glasgow.ac.uk}{daniele.faccio@glasgow.ac.uk}}

\begin{abstract}
\noindent{}Fluorescence lifetime imaging microscopy (FLIM) provides detailed information about molecular interactions and biological processes. A major bottleneck for FLIM is image resolution at high acquisition speeds, due to the engineering and signal-processing limitations of time-resolved imaging technology. Here we present single-sample image-fusion upsampling (SiSIFUS), a data-fusion approach to computational FLIM super-resolution that combines measurements from a low-resolution time-resolved detector (that measures photon arrival time) and a high-resolution camera (that measures intensity only). To solve this otherwise ill-posed inverse retrieval problem, we introduce statistically informed priors that encode local and global dependencies between the two “single-sample” measurements. This bypasses the risk of out-of-distribution hallucination as in traditional data-driven approaches and delivers enhanced images compared for example to standard bilinear interpolation. The general approach laid out by SiSIFUS can be applied to other image super-resolution problems where two different datasets are available. 
\end{abstract}

\maketitle


\clearpage 
\section{Introduction}
\noindent{}Fluorescence lifetime imaging microscopy (FLIM) finds extensive applications in biological studies where the lifetimes of fluorophores can be used as indicators of the cellular metabolism \cite{georgakoudi2012optical, stringari2015vivo, stringari2017multicolor, yaseen2017fluorescence, schaefer2019nadh}, cellular environment \cite{okabe2012intracellular, ogikubo2011intracellular, suhling2019fluorescence} or changes in molecular conformation visible through Förster resonance energy transfer (FRET), enabling measurement of protein:protein interactions during processes such as cellular signalling \cite{van2005fluorescence, bajar2016guide, datta2020fluorescence, berezin2010fluorescence}. In medical settings, endogenous FLIM can be used for identifying cancerous tissue \cite{sun2010fluorescence, gorpas2019autofluorescence}. \\
FLIM setups excite a sample with short-wavelength light and measure the temporal profile of long-wavelength fluorescence from the sample \cite{lakowiczTimeDomainLifetimeMeasurements2006}. Excitation is achieved using a pulsed or amplitude modulated laser for timedomain and frequency domain FLIM, respectively \cite{datta2020fluorescence}, while emission is usually collected with time correlated single photon counting (TCSPC) or time-gated hardware. Fluorescence lifetime is then recovered from the temporal decay of fluorescence emission. Popular lifetime estimation schemes include least squares deconvolution \cite{lakowicz2006principles}, Laguerre expansion \cite{jo2005ultrafast}, phasor fitting \cite{stringari2015vivo,stringari2017multicolor}, rapid lifetime determination \cite{ballew1989error, elson2004real}, centre-of-mass estimation \cite{poland2016new, laine2022method} and machine learning \cite{smith2019fast,zickus2020fluorescence,kapitany2023single}.\\
Images are formed through raster-scanning or widefield detection. Scanning systems allow confocal or 2-photon microscopy setups, giving excellent image resolution, and aligning well with TCSPC methods that give rich fluorescence information. However, scanning also presents drawbacks, such as the lack of instantaneous complete field-of-view information, and long acquisition times which are incompatible with the rapid intracellular dynamics of living cells \cite{knopfel2019optical, dvinskikh2023high}. Widefield systems overcome these challenges by measuring temporal decay from the full field of view in parallel, often using time-gated cameras like intensified charge-coupled devices (iCCDs) \cite{mcginty2010wide,kelly2014automated}, externally gated devices \cite{bowman2019electro,bowman2023wide}, or single photon avalanche diode (SPAD) arrays \cite{zickus2020fluorescence, bruschini2019single}. However, iCCD resolution is limited by the intensifier point-spread-function, whilst SPAD arrays typically have low-pixel counts and/or low fill-factors.\\
Computational super-resolution (SR) provides a route to overcome the trade-off between acquisition time and spatial resolution by offloading imaging from optics onto software. SR takes an undersampled image of a scene and estimates its high-resolution features. Multiple flavours of SR exist, which are generally either interpolation, reconstruction (inverse retrieval) or example- (learning) based.\\
Interpolation is the simplest form of upsampling, encompassing several methods for connecting datapoints with some curve \cite{matplotlib}. For images, this ranges from simple schemes like nearest, bilinear and bicubic interpolation, through frequency-based approaches sinc and Lanczos interpolation, to covariance-based algorithms like kriging (Gaussian processes) \cite{fadnavis2014image}. While interpolation is fast and computationally inexpensive, it does not add new information to the image.\\
Reconstruction-based modeling instead manipulates the detection to optically redistribute information about the high-resolution target into fewer measurements. This encoding provides a mathematical forward model that is employed to reconstruct the non-sampled points in an inverse retrieval framework, for example via point spread function (PSF) engineering \cite{sun2020end}, blurring \cite{callenberg2021super}, or compressed sensing \cite{candes2006robust,sun2018depth, soldevila2021giga, antipa2018diffusercam}.\\
Lastly, example-based schemes rely on computation and pattern recognition to upsample images in a data driven manner \cite{freeman2002example}. Classical approaches include neighbour embedding \cite{chang2004super}, sparse coding \cite{yang2010image} and anchored neighbourhood regression \cite{timofte2013anchored}. More recently, machine learning algorithms have seen widespread adoption for super-resolution \cite{wang2020deep,kapitany2023ai}. These range from super-resolution convolution neural networks \cite{dong2014learning,dong2016accelerating}, through generative adversarial networks \cite{ledig2017photo,wang2018esrgan}, to diffusion models \cite{saharia2022image}. However, learning-based schemes traditionally need large, diverse training datasets, which can pose a bottleneck in niche fields like FLIM; further, different fluorophores behave differently, hampering generalisation in traditional machine learning methods \cite{mannam2020machine}. Self-similarity-based super-resolution \cite{glasner2009super} and self-supervised clustering \cite{quiros2022self} approaches offer an alternative to external training set, deriving statistical information for super-resolution from the very image that is up-sampled.\\
Data from different sensing modalities can yield more information about a subject than is contained in each modality alone \cite{kapitany2023ai,huang2020fusion}. Fusion-based inference is a growing field with applications from medical imaging using PET and MRI \cite{thung2017multi}, through autonomous driving using camera and LiDAR \cite{person2019multimodal}, to content classification using video and text \cite{trzcinski2018multimodal}. Data fusion has been applied to FLIM, by interpolating lifetime images and weighting them with intensity images for visualisation \cite{kapitany2023single,samimi2023light}.\\
Here we introduce a super-resolution method that relies on the fusion of two images: a high-resolution intensity image (no lifetime information) and a low-resolution lifetime image. Our method is called ‘single sample image fusion upsampling’ (SiSIFUS). SiSIFUS generates data-driven lifetime priors matching the resolution of the intensity image; this is relatively easy and inexpensive to acquire at high-resolution, compared to FLIM images.\\
Crucially, our method generates ‘single sample’ priors: all information in our scheme comes from the given field of view, not external training data. We develop two priors, which extract this information from the FLIM-intensity image pair in different ways. Local priors correlate low resolution FLIM pixels with corresponding intensity pixels in small neighbourhoods. Global priors instead exploit morphological signatures in the image, using a neural network to predict fluorescence lifetime from intensity patches.\\
SiSIFUS combines data fusion and self-supervised learning into a practical super-resolution framework. Like example-based self-similarity approaches, it avoids complex hardware modifications and external training data. Like reconstruction-based modelling, we optically measure high-resolution features, giving more information than is available in the low- resolution images alone.\\
\section{Results}
\begin{figure*}[t]
    \centering
    \includegraphics[width=\linewidth]{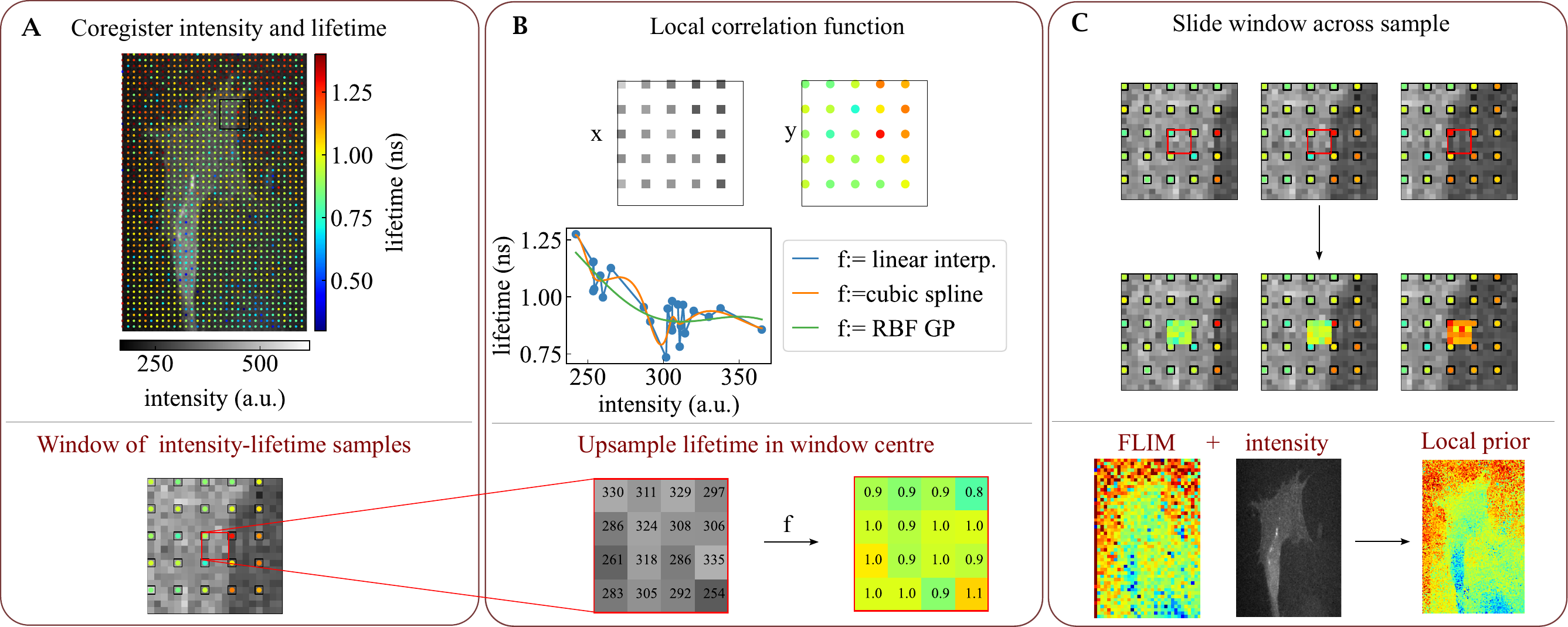}
    \caption{\textbf{Schematic of the local prior method.} \textbf{(A)} Shown are a CMOS (fluoresence intensity) field of view, with the SPAD field of view (fluorescence lifetime), overlayed on top of it so as to match the sparse, low fill-factor pixel layout of the SPAD array. \textbf{(B)} We zoom in on a 5 × 5 window. All SPAD pixels have a corresponding CMOS measurement, but so do the areas in-between SPAD pixels. We aim to find the lifetime at points with no SPAD samples. For this, we fit a function, for instance linear interpolation, a cubic spline or a radial basis function gaussian process. Then, the high-resolution CMOS pixels $x_{HR}$ which we wish to upsample are fitted with this function, producing a lifetime estimate $\hat{tau}_{HR}$. (C) We slide the window across the field of view, fitting new functions for each new window, and predicting the centres, upsampling the FLIM image to the resolution of the intensity image, window-by-window.}
    \label{fig:1}
\end{figure*}
\noindent{\bf{Forward and inverse models.}} We apply SiSIFUS to both raster-scanning and widefield FLIM. The scanning system uses a PMT to gather both the FLIM and intensity image, while the widefield system uses a SPAD array to measure FLIM, and a complementary metal–oxide–semiconductor (CMOS) camera to measure intensity. Both of the setups are detailed in the Methods.\\
SiSIFUS involves two measurements. The first is the time-resolved, low-resolution datacube, $r\in{\mathbb{N}^{m,n,t}}$, where $m,n$ denote spatial position and $t$ denotes time. The fluorescence lifetime image, $\tau_{LR}\in{\mathbb{R}^{m,n}}$, is estimated from $r$ via a standard least squares deconvolution - other schemes, like phasor analysis or centre-of-mass estimation, could be used equivalently.\\
The second measurement is the high-spatial resolution intensity measurement, $I\in{N^{M,N}}$, where $M,N$ denote the pixel numbers of the high-spatial resolution sensor. SiSIFUS then super-resolves the lifetime image $\tau_{LR}$ to match the pixel count of the intensity image $I$.\\
Our setups sample fluorescence lifetime sparsely across the field of view. In the widefield setup, this arises from low fill-factor, ergo large dead spaces between the active areas of the SPAD pixels. In the scanning setup, this arises from the large sampling period relative to the spot size of the excitation beam in the object plane. We also assume that the intensity measurement has approx. 100\% fill-factor. Fig. \ref{fig:1}A and Fig. \ref{fig:2}A depict how intensity and lifetime are sampled.\\
For 256×256 sized high-resolution intensity image $I$ of the sample, the acquired dataset is integrated along the time axis. In a practical scenario for upsampling a confocal scan image, the FLIM samples would be acquired by taking a large line-average of low-resolution scans. A large line-average is needed for the fitted lifetime to have decent signal-to-noise ratio (SNR). The intensity image has decent SNR even with just a few line-averages, therefore the high-resolution intensity image could be obtained without adding an external sensor, by simply scanning a second time, with a higher pixel count but much fewer line averaging.\\
Consequently, $\tau_{LR}$ is decimated (sparsely sampled) from the high-resolution fluorescence lifetime target, $\tau_{HR}\in{\mathbb{R}^{M,N}}$, that we aim to reconstruct:	
\begin{equation}\label{eq:1}
    \tau_{LR} = \mathbf{A}\tau_{HR}
\end{equation}
where $\mathbf{A}$ represents sparse sampling (decimation).\\
We feed the two images, $I(M,N)$ and $\tau_{LR}(m,n)$, to our prior-generation pipeline (explained below), which outputs a local and global prior, $\hat{\tau}_{LP}(M,N)$ and and $\hat{\tau}_{GP}(M,N)$, respectively. These priors constrain an (otherwise ill-posed) inverse retrieval algorithm. We finally recover the high-resolution lifetime image $\tau_{HR}{*}$ by minimizing the following cost function: \\
\begin{equation}\label{eq:2}
\begin{aligned}
\begin{split}
    \tau_{HR}{*} = & \underset{\hat{\tau}_{HR}}{\arg\min}\;C(\hat{\tau}_{HR}) \text{, where}\\
    C(\hat{\tau}_{HR}) &=\|\mathbf{A}\hat{\tau}_{HR}-\tau_{LR}\|_2^2+ \gamma\|\hat{\tau}_{HR}-\hat{\tau}_{LP}\|_2^2\\
    & + \beta\|\hat{\tau}_{HR}-\tau_{GP}\|_2^2 +\alpha\|{\mathbf{D} \hat{\tau}_{HR}}\|_1\\
    & \text{subject to } \hat{\tau}_{HR}\geq 0
\end{split}
\end{aligned}
\end{equation}
The first term in $C(\hat{\tau}_{HR})$ ensures the data fidelity between the low-resolution measured lifetime image and the downsampled optimal high resolution lifetime solution in each iteration. Prior constraints on the target high resolution lifetime image are enforced through the second and the third data fidelity term, weighed by the factor $\gamma$ and $\beta$ respectively, which are empirically optimised to yield best results. The fourth term is the L1-norm of the 2D total variation (TV) evaluated on the high-resolution lifetime image and weighed by $\alpha$ \cite{ji2022compressed}. We consider the anisotropic form of the TV \cite{chambolle2004algorithm}, and so the operator $\mathbf{D}$ represents the finite differences approximation of the horizontal and vertical image gradients.
\subsection{Dependence between lifetime and intensity.}
\noindent{}SiSIFUS priors exploit inter-dependence between fluorescence lifetime and intensity. Although these variables are interdependent at the single molecule level via fluorescence quantum yield, this dependence is modulated by fluorophore concentration and other complex and often unpredictable biophysical mechanisms, thus necessitating statistical methods to create our priors.
\begin{figure*}[t]
    \centering
    \includegraphics[width=\linewidth]{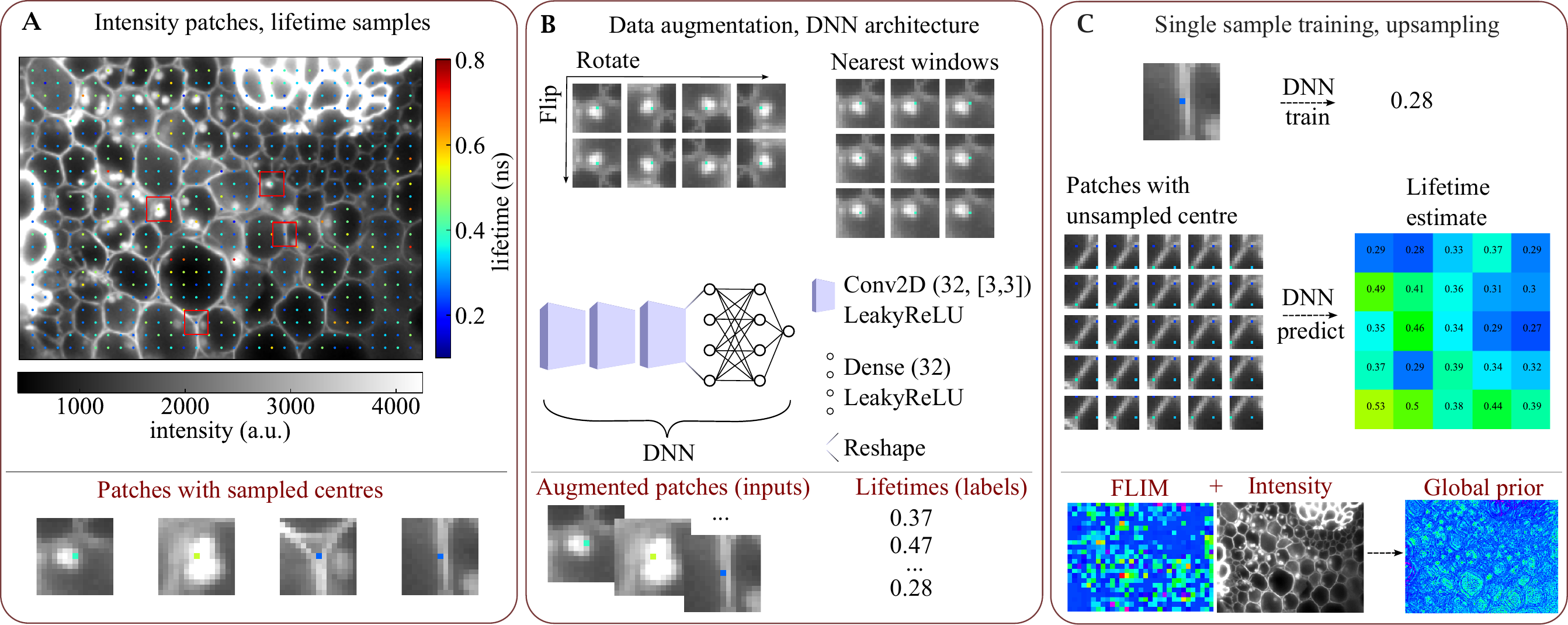}
    \caption{\textbf{Schematic of the global prior method.} \textbf{(A)} Fluorescence intensity of a convallaria - acridine orange sample, with 8 × 8 sparse lifetime samples overlayed. We extract intensity patches from this image; a few of them correspond to a central lifetime sample. Such patches are training data, which we can use to predict the central lifetime of the rest of the patches. \textbf{(B)} Training inputs (patches) are augmented via rotation and mirroring. They can be further augmented by adding the patches which are nearest neighbours of training patches and allocating them the same label (lifetime) as the sampled patch. The deep neural network (DNN) architecture is simple, consisting of three 2D convolutional layers followed by three fully connected layers. \textbf{(C)} Finally, the trained DNN evaluates patches with unsampled centres, thus super-resolving the lifetime image.}
    \label{fig:2}
\end{figure*}\\
Fluorescence quantum yield $Q$ is the ratio of the number of emitted photons to the number absorbed. It depends on the radiative and non-radiative decay rates $k_r$ and $k_nr$ that depopulate excited molecules. The measured fluorescence lifetime $\tau$ also depends on these rates \cite{lakowicz2006principles}:
\begin{align}\label{eq:3}
    Q &= \frac{k_r}{k_r+k_{nr}}\nonumber \\
    \tau &= \frac{1}{k_r+k_{nr}}
\end{align}
therefore $Q=k_r\tau$ for a single molecule. Across a given field-of-view, fluorescence intensity variations are given by the quantum yield of fluorophores (equivalently, fluorescence lifetime) multiplied by their absorbance (concentration times absorptivity times sample thickness). Absorbance is typically unknown and unpredictable, hence it acts as a confounding variable in intensity-lifetime dependencies, so fluorescence intensity alone cannot give us full lifetime information. This means that two samples might have the same lifetime, but completely different intensities, or vice versa.\\
However, across a single sample, fluorophore concentration typically varies slowly compared to lifetime and/or covaries with it on local scales, such that it is possible to build local priors that capture the resulting intensity-lifetime dependencies. Further, absorbance and lifetime often co-vary with cellular morphology, enabling us to create global priors. As a fail-safe, if neither local nor global dependencies exist across a specific sample or sub-region, TV-minimisation (a form of edge-preserving interpolation) in our inverse retrieval ensures that our method still performs at least as well as standard interpolation (see supplemental material for details).\\
\noindent{\bf{Local prior.}} The local prior (LP) relies on direct, pixelwise dependencies between lifetime and intensity on micron scales: Fig. \ref{fig:1} illustrates our workflow. If the images come from different detectors, the lifetime and intensity image are first coregistered to match their fields of view. Fig. \ref{fig:1}A shows a sparse, low-resolution lifetime image (RGB) overlayed on the corresponding intensity image (grayscale). The field of view (FOV) is divided into windows, each containing a set of corresponding intensity-lifetime samples. These samples neighbour intensity pixels in the window centre, hence this window is used to create a prior for those pixels. In each window, the intensity and lifetime pairs are vectorised and fitted with a function, $f$ - see Fig. \ref{fig:1}B. Thus, our lifetime estimate $\hat{\tau}$ for pixel $(\lambda i+x,\lambda j+y)$, is:
\begin{equation}\label{eq:4}
    \hat{\tau}_{\lambda i + x, \lambda j + y} = f_{i,j}(I_{\lambda i + x, \lambda j + y})
\end{equation}
with samples $i\in{\{0,1,...,n-1\}}$ and $j\in{\{0,1,...,m-1\}}$, and $x\geq 0,y< \lambda$. Importantly, the functions $f_{i,j}$ are fitted \textit{locally}, not globally. Consequently, this procedure is repeated by sliding the window across the field of view, as shown in Fig. \ref{fig:1}C.\\
\noindent{\textbf{Global prior}}. Images often contain multiple examples of similar features, with similar lifetime distributions, across the field of view. This motivates our development of global priors (GPs) that exploit correlations between high-resolution morphology and lifetime.\\
Fig. \ref{fig:2} shows our pipeline. We first extract intensity patches centred on our SPAD pixels, as shown in Fig. \ref{fig:2}A. To deal with the relatively small number of patch-lifetime pairs contained in a single sample image, we augment our training set. We use a commonly used dataset augmentation technique by reflecting and rotating the intensity windows in the training set. These operations increase our dataset 8-fold, as shown in Fig. \ref{fig:2}B. For high upsampling factors (8 × 8 and 16×16), we further augment the training set by estimating the lifetimes of the patches neighbouring our sampled patches. We then label these patches with the same lifetime as their sampled neighbour. Our approach is visualised in Fig. \ref{fig:2}B; see Methods for details.\\
Our global priors are designed to generalise to new samples with previously unseen morphologies, morphology-lifetime dependencies, and lifetime ranges. A deep neural network (DNN), shown in Fig. \ref{fig:2}B, is trained from scratch for each new sample, on the intensity-patch inputs and lifetime labels obtained from the given microscope field of view. Consequently, different DNN initialisations give slightly different predictions. To estimate high-resolution lifetime, we pass each intensity patch through our trained DNN, predicting the central lifetime value, as shown in Fig. \ref{fig:2}C.\\ 
\noindent{\bf{Quality metrics.}} We track reconstruction quality using three metrics: learned perceptual image patch similarity (LPIPS), structural similarity index measure (SSIM) and peak signal-to-noise ratio (PSNR). LPIPS measures distance between images in feature space. It has a minimum of 0 and grows with image dissimilarity (higher values are worse). SSIM tracks the similarity in luminance, contrast and structure between two images; it is bound between -1 and 1, with larger values indicating better image similarity. Lastly, PSNR is a pixel-to-pixel comparison, where larger values are better. See the Supplement for details.
\begin{figure*}
    \centering
    \includegraphics[width=\linewidth]{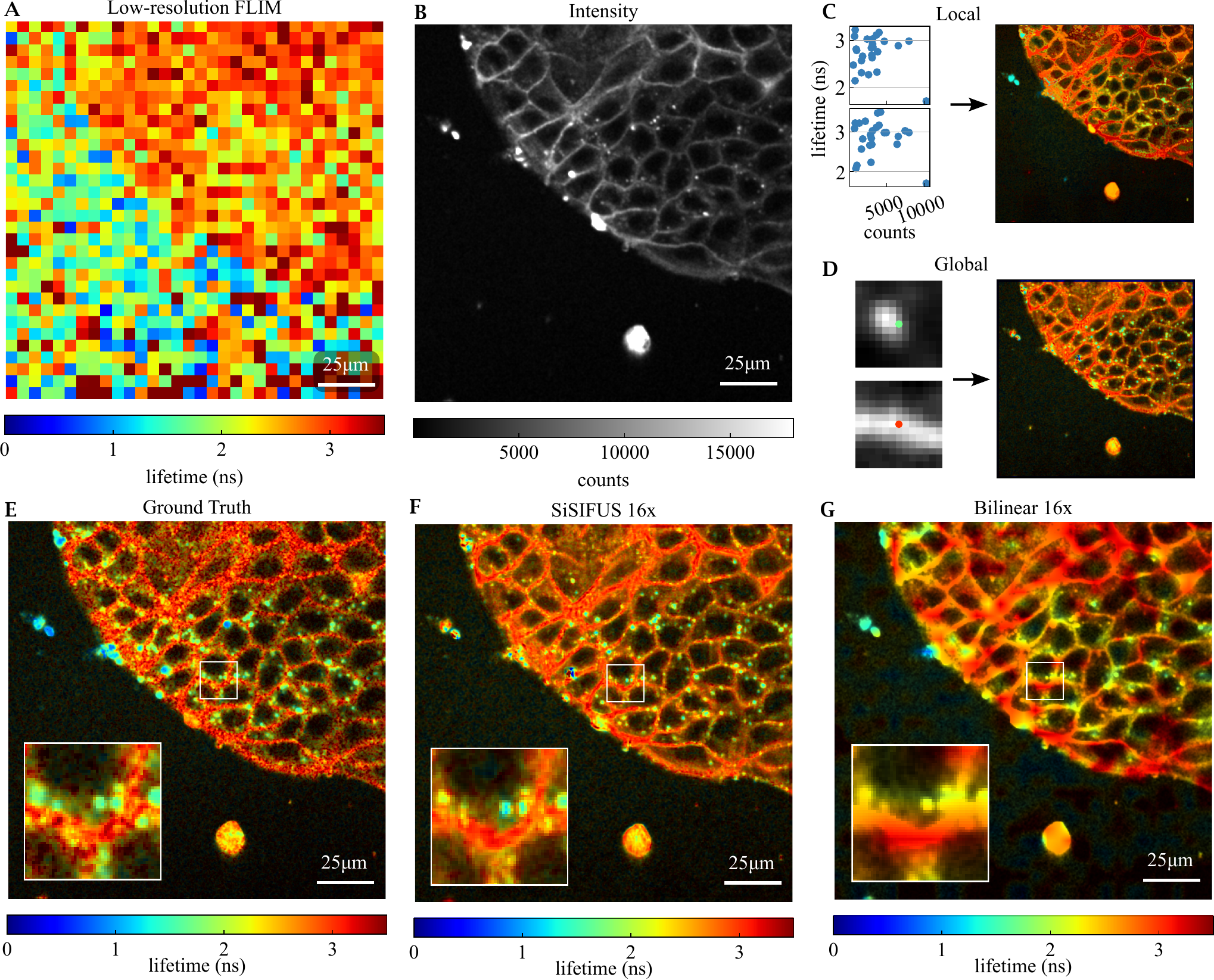}
    \caption{\textbf{16x16 upsampling of MDCK cells.} \textbf{(A)} Low resolution fluorescence lifetime image (32x32) of Madin-Darby canine kidney (MDCK) cells expressing Flipper-TR dye. \textbf{(B)} Corresponding high resolution intensity image (512x512) of the sample. \textbf{(C)} 5x5 windows of low-resolution FLIM are fitted to corresponding intensity values, to generate a local prior image (two example windows are shown). \textbf{(D)} A global prior image is generated from 13x13 intensity patches with central FLIM measurements (two examples are shown). \textbf{(E)} The ground truth high-resolution FLIM target, intensity weighted for visualisation. \textbf{(F)} The proposed method, upsampling the low-resolution measurement by a factor of 16x16. \textbf{(G)} Bilinear interpolation upsampling the FLIM measurement by 16x16.}
    \label{fig:3}
\end{figure*}\\
\noindent{\bf{Sample 1: 16x16 (MDCK Flipper-TR).}} We examined a validation sample of Madin-Darby canine kidney (MDCK) cells that had been treated with Flipper-TR dye (Spirochrome Inc.), which allows quantitation of tension in living, migrating cells. Data was acquired using a commercial LaVision BioTec TriM Scope system, using two-photon excitation scanning, and detecting emission via a photo-multiplier tube (PMT). The sample was imaged at 512 × 512 spatial points covering a 167 × 167 $\mu m^2$ FOV, and binned into 75 time bins, giving a 512 × 512 × 75 datacube. See Methods for details.\\
Ground truth (GT) fluorescence lifetime was estimated from this datacube using least squares deconvolution. This was decimated 16-fold to give a low-resolution FLIM image, shown in Fig. \ref{fig:3}A. The low-resolution FLIM image is severely undersampled; a lot of the detail was lost. Fluorescence intensity (Fig. \ref{fig:3}B) was obtained in parallel, by summing the datacube along time. In the intensity image, we see that the probe mainly localised to two types of structures: small blobs (vesicles) and edges (cell membranes). In Fig. \ref{fig:3}(C-D) we show 3 examples of local prior windows and corresponding global prior patches, extracted as shown in Fig. \ref{fig:1}B and Fig. \ref{fig:2}A, respectively. Local (pixelwise) dependencies appear relatively weak, instead the global (morphological) dependencies dominate, capturing the trend that vesicles have lower lifetimes than cell membranes across the FOV.\\
Fig. \ref{fig:3}E shows the ground truth lifetime, weighted with local contrast enhanced fluorescence intensity for visualisation (details in the Supplement). In Fig. \ref{fig:3}(F-G) we show SiSIFUS and bilinear interpolation for the upsampling task. SiSIFUS automatically learns to distinguish between vesicles and cell membranes and labels them with different lifetimes, whereas interpolation fails to reconstruct fine details. SiSIFUS also maintains sharp boundaries between structures of different lifetimes, informed by the intensity image. SiSIFUS has LPIPS of 0.24, SSIM of 0.31, and PSNR of 26dB. Interpolation, instead, has LPIPS of 0.48, SSIM of 0.12, and PSNR of 24dB.%
\begin{figure*}
    \centering
    \includegraphics[width=\linewidth]{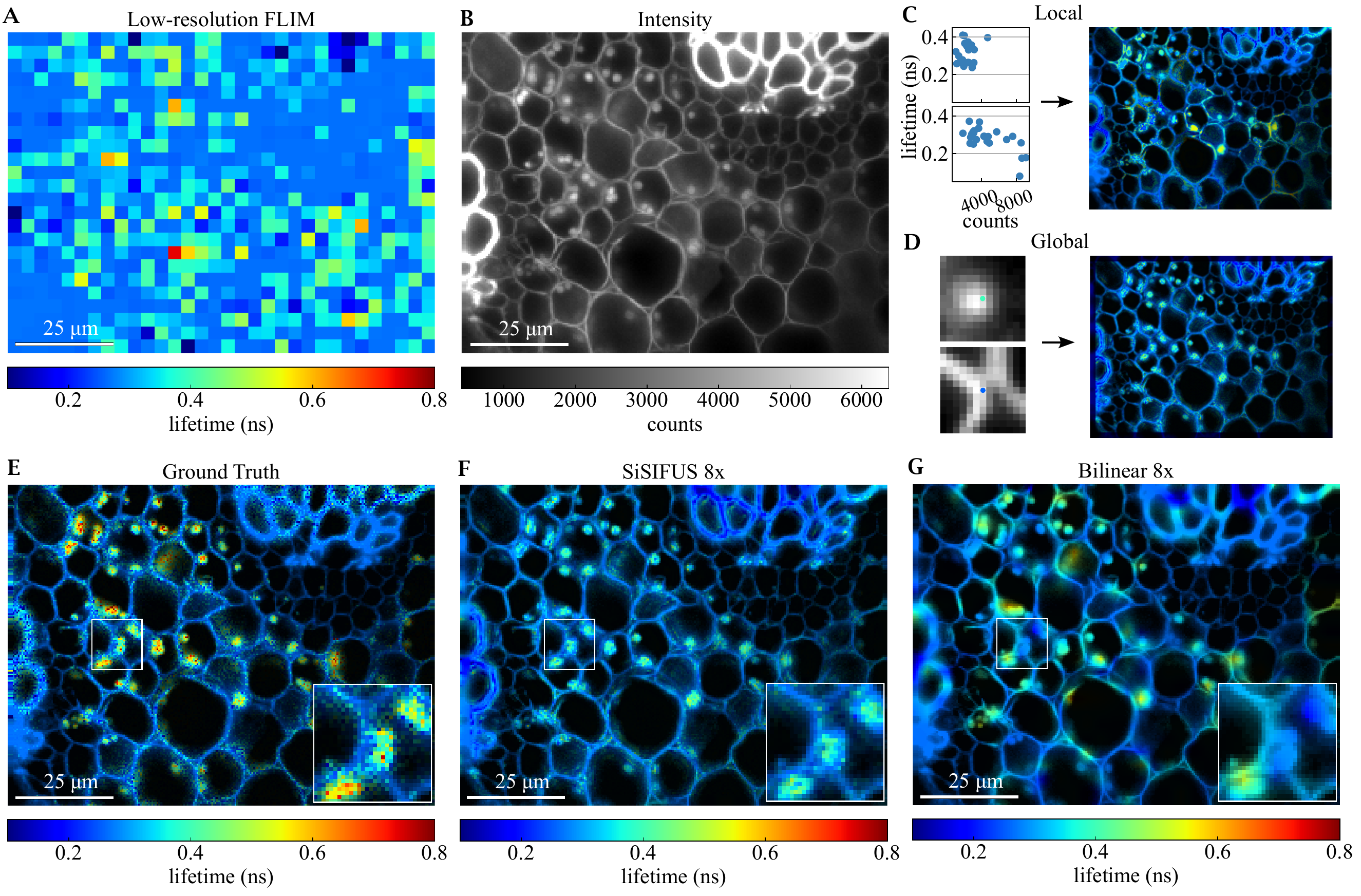}
    \caption{\textbf{8x8 upsampling of convallaria images.} \textbf{(A)} Low-resolution fluorescence lifetime image (24 × 32) of a convallaria rhizome sample stained with Acridine Orange, viewed under a widefield microscope. \textbf{(B)} High-resolution intensity image (192 × 256). \textbf{(C)} Example 5x5 windows of low-resolution intensity vs FLIM, used for generating the local prior shown on the right. \textbf{(D)} High-resolution intensity patches are labelled with lifetime, letting us create a global prior. \textbf{(E-G)} Ground truth, 8x8 SiSIFUS and 8x8 bilinear interpolation of the data, weighted by local contrast enhanced intensity for visualisation – see Supplementary Materials for details.}
    \label{fig:4}
\end{figure*}\\
\noindent{\bf{Sample 2: 8x8 (Convallaria AO).}} We applied SiSIFUS to a Convallaria rhizome sample dyed with Acridine Orange. The fluorescence lifetime datacube was recorded using our custom microscope setup, which uses a FLIMera 192×128-pixel SPAD array (HORIBA Scientific \cite{henderson2019192}) to image an 82×107$\mu m^2$ FOV, using 326 time bins. To compensate for the HORIBA camera’s asymmetric pixel layout, we scanned the sample twice with a pixel shift, giving a 192×256×326 datacube. Simultaneously, a high spatial resolution sCMOS camera registered a 2048×2048-pixel image of the sample. The sCMOS camera is spatially co-registered to match the SPAD’s field of view (FOV) and resolution (see Methods).\\
The low-resolution FLIM and high-resolution intensity guide are shown in Fig. \ref{fig:4}(A-B). Fig. \ref{fig:4}(C-D) shows example local and global dependencies and priors, respectively. Global dependencies appear to dominate local ones for this sample, with globules having shorter lifetimes than cell walls.\\
The 8×8 upsampling results in Fig. \ref{fig:4}(E-G) illustrate that SiSIFUS recognises that globules tend to have higher lifetimes than cell walls, hence maintaining contrast between these structures more consistently than bilinear interpolation. We do note though that global SiSIFUS misses certain hotspots in the GT lifetime image (high-lifetime, yellow/red coloured areas), likely because few globules in the training set have these lifetimes, hence the model treats them as outliers. SiSIFUS achieves an LPIPS of 0.11, SSIM of 0.21, PSNR of 16dB. Bilinear interpolation has an LPIPS of 0.15, SSIM of 0.29, and PSNR of 16dB.
\begin{figure*}
    \centering
    \includegraphics[width=0.94\linewidth]{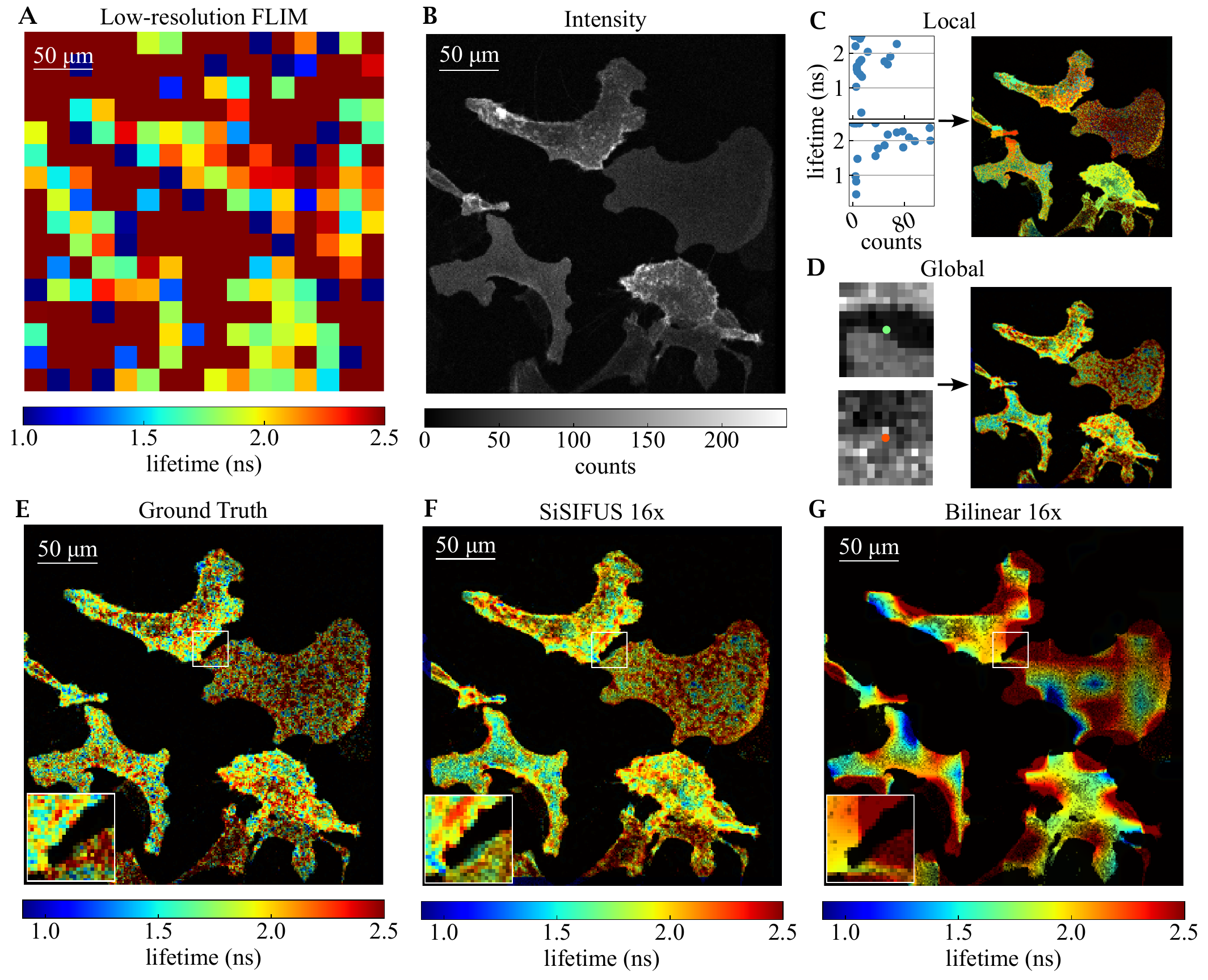}
    \caption{\textbf{16x16 upsampling of SKOV3 cells.} \textbf{(A)} Low-resolution fluorescence lifetime image (16×16) of a SKOV3 samples expressing Rac1-Raichu. \textbf{(B)} High-resolution intensity image (256×256). \textbf{(C-D)} Examples of local and global dependencies. \textbf{(E-F)} Ground truth, 16×16 super-resolved, and 16×16 interpolated images.}
    \label{fig:5}
\end{figure*}\\
\textbf{Sample 3: 16x16 (SKOV3 - Rac1 Raichu)}. We further validate SiSIFUS on measurements of SKOV3 ovarian cancer cell samples expressing Rac-Raichu Clover mcherry (see Methods). The GT images are acquired using our LaVision BioTec TriM Scope two-photon scanning system. The field of view was sampled on a 256×256 square grid covering 301×301 $\mu m^2$ area. The temporal evolution was recorded using TCSPC with 75 timebins of 160ps duration each, giving a fluorescence datacube of size 256×256×75.\\
The low-resolution fluorescence lifetime input is shown in Fig. \ref{fig:5}A, alongside the high-resolution intensity guide in Fig. \ref{fig:5}B. Fig. \ref{fig:5}(C - D) show a set of local windows and the LP, as well as global patches and the GP, respectively.\\
Local dependencies exhibit plateauing positive correlations and seem to flatten fluorescence lifetimes across the different cells, capturing inter-cellular dynamics. Global priors instead capture intra-cellular dynamics, showing that fluorescence lifetime is mostly uniform within cells, with some patterned textures. Since the upsampling factor is high (16x16), our algorithm prioritises global priors.\\
Fig. \ref{fig:5}(E - G) show the ground truth compared to 16x16 SiSIFUS and bilinear interpolation; SiSIFUS gives a better estimate. SiSIFUS reconstruction has an LPIPS to the ground truth of 0.15, SSIM of 0.08, and 12dB PSNR. In contrast, interpolation has an LPIPS of 0.34, SSIM of 0.06 and PSNR of 11dB.
\begin{figure*}
    \centering
    \includegraphics[width=0.75\linewidth]{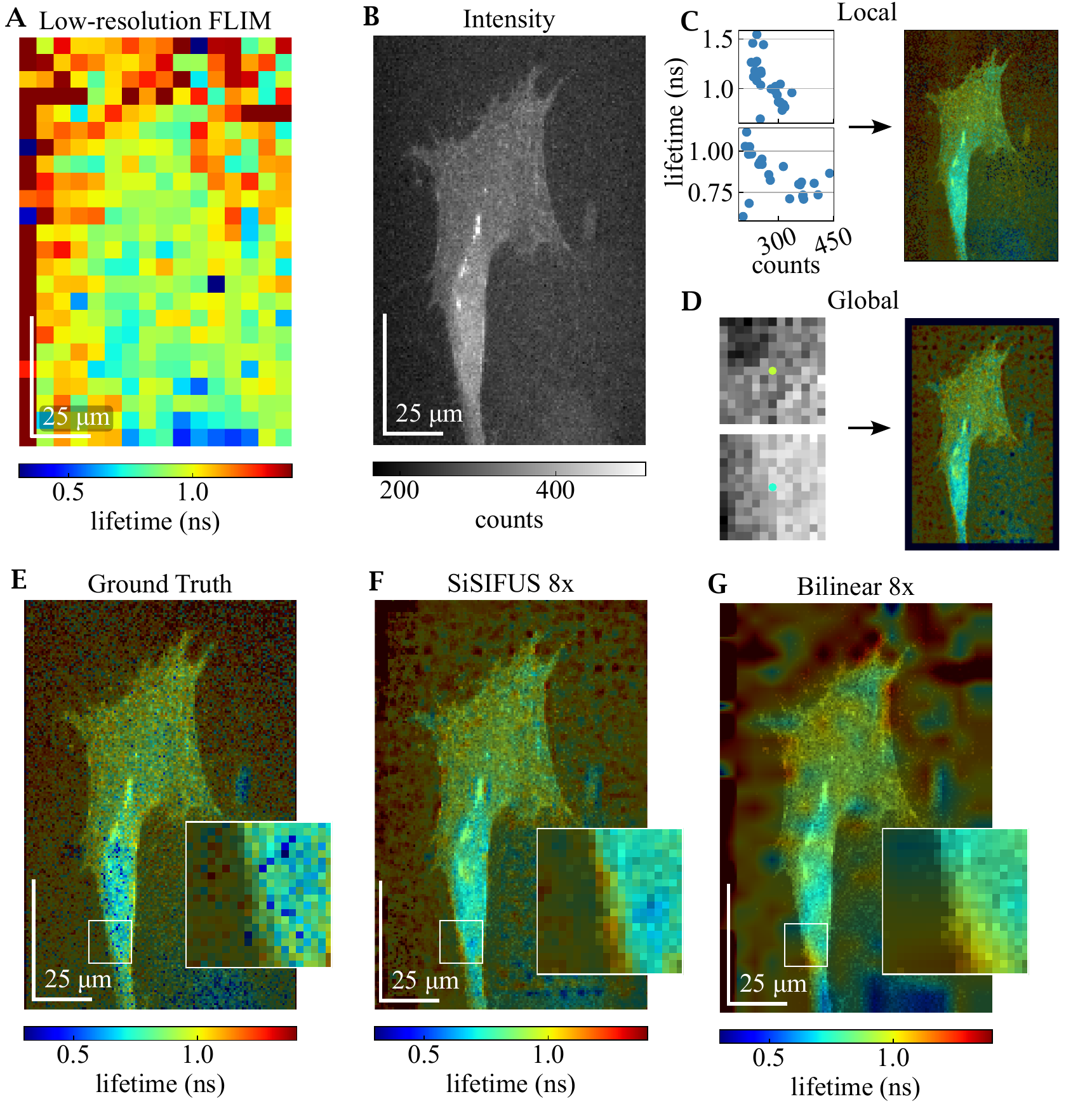}
    \caption{\textbf{8x8 super-resolution of SKOV3 images.} \textbf{(A)} Low-resolution FLIM image of a SKOV3 cell expressing the Rac1-Raichu probe (24×16). \textbf{(B)} The corresponding high-resolution fluorescence intensity image (192×128). \textbf{(C-D)} Comparison of local and global intensity-lifetime dependencies observed in this sample, and the corresponding local and global prior images. \textbf{(E-G)} High-resolution ground truth FLIM, 8×8 SiSIFUS and 8×8 bilinearly interpolated images, respectively.}
    \label{fig:6}
\end{figure*}\\
\noindent{\bf{Sample 4: 8x8 (SKOV3 Rac1-Raichu).}} We applied SiSIFUS to widefield images of SKOV3 ovarian cancer cells expressing Rac-Raichu Clover mcherry, acquired with our custom FLIMera SPAD array and Zyla sCMOS setup (see Methods). The temporal evolution was recorded using TCSPC with 326 timebins, giving a fluorescence datacube of size 192 × 128 × 326. This image was decimated to give the low-resolution FLIM input.\\
Fig. \ref{fig:6} shows our results. The low-resolution FLIM (Fig. \ref{fig:6}A) and high-resolution intensity (Fig. \ref{fig:6}B) are used to generate local and global priors, which are then fed into our inverse retrieval algorithm. Fig. \ref{fig:6}C and Fig. \ref{fig:6}D shows examples of local and global intensity lifetime dependencies and priors.\\
This sample shows non-linear negative local interdependencies at most regions; SiSIFUS can exploit these to accurately determine the lifetime based on local intensity patterns. Conversely, global patch-lifetime dependencies are negligible. This is mainly because the field of view lacks repeating morphological features (in contrast to the MDCK and convallaria samples in Fig. \ref{fig:3} and Fig. \ref{fig:4}). Our algorithm prioritises the local priors.\\
Finally, Fig. \ref{fig:6}(E-G) shows the ground truth compared to SiSIFUS and bilinear interpolation. SiSIFUS succeeds in reconstructing the lifetime boundaries seen at the cell edges, and also reconstructs the speckliness of the ground truth lifetime map, allowing the user to infer that there might be lifetime estimation uncertainty. Interpolation fails in these regards: edges are blurred, and lifetime estimates appear smooth give the impression of structures that are absent in the ground truth. SiSIFUS yields an LPIPS of 0.31, SSIM of 0.21, and PSNR of 16dB, whereas interpolation has an LPIPS of 0.56, SSIM of 0.22, and PSNR of 16dB.
\subsection{Acquisition times}
\noindent{}SiSIFUS provides an advantage in terms of acquisition times. For example, if we consider the case of measurements taken with our TriM Scope I (Figures \ref{fig:3} and \ref{fig:5}), the acquisition time scales linearly with pixel number as this is a galvo-scanning system. Therefore, we have an immediate advantage given by the SiSIFUS resolution enhancement factor that is applied. Specifically, in Figure 3, where we apply 16x16 resolution enhancement, we have a 256x reduction in the number of points that need to be scanned and hence a 256x reduction in acquisition time. In a scanning system we still, however, need to perform a second scan for the high-resolution intensity measurement but this typically can be at substantially higher speed, of order 35x in our system (and this therefore remains the limiting factor). If we therefore consider the specific case of a 512x512 image (Figure 3), the total acquisition time without SiSIFUS was 73 seconds and with SiSIFUS is 2.4 seconds allowing a 0.4 fps acquisition rate.\\
If instead we consider the case of measurements with a SPAD camera (in our case, the Horiba FLIMera system, Figures \ref{fig:4} and \ref{fig:6}) this currently operates at 30 fps, i.e. 33 ms acquisition time for a SiSIFUS image of any size (all pixels are acquired in parallel without any point-by-point scanning used in confocal imaging systems). We note that in this case, the intensity CMOS image is acquired in parallel and hence does not add to the acquisition time.\\
We may compare this also with existing commercial systems, e.g. current B\&H FLIM systems can measure 512x512 pixels in 1 second \cite{becker2023spc} or previous work that operated directly with megapixel SPAD arrays in which, however, the smaller pixel size implied longer acquisition times in order to accumulate sufficient signal and was thus limited to $\sim$1 second acquisition times \cite{zickus2020fluorescence}.
\section{Discussion}
\noindent{}We introduce SiSIFUS, a robust, data-fusion pipeline based on prior-augmented inverse retrieval for upsampling fluorescence lifetime images. We create two classes of priors that explicitly exploit a high-resolution intensity image to provide approximations for the non-sampled datapoints in a fluorescence lifetime image. The goal of SiSIFUS is to provide a “physics inspired” approach to image resolution enhancement that performs better than standard bilinear or similar interpolation methods.\\
Local priors capture pixel-wise correlations between fluorescence lifetime and intensity. For this, we find a direct mapping from intensity values to lifetime in small, local neighbourhoods, and use this mapping to predict the lifetime of intensity pixels that lack corresponding lifetime pixels. This allows SiSIFUS to maintain sharp spatial boundaries, tracking the boundaries of our intensity image. The local prior is limited by measurement noise and sampling frequency. Since structures of similar intensity are assigned the same lifetime, undersampled regions may receive homogeneous lifetime estimates with sharp boundaries, as seen in the leftmost cell in Fig. \ref{fig:5} (c). Noisy regions can instead artificially track the intensity of an image's noise, as in Fig. \ref{fig:6}(c). TV-minimization and the global prior help combat these issues.\\
Global priors capture inter-dependence between FLIM and intensity on a morphological level. This is achieved by learning a mapping (deep neural network) from intensity patches to central-pixel lifetime samples, and then using mapping to predict the lifetime of patches that have no central lifetime measurements. Thus, we capture non-linear correlations between the brightness and shape of intensity features and lifetime. In microscopic samples, there often exist strong global trends between these variables, allowing the model to predict the lifetimes of patches with unsampled centres. However, the global prior has limited ability to distinguish between similar morphologies with different lifetimes, typically assigning them with the average lifetime of such structures. This causes outliers like the low-lifetime globules on the left of Fig. \ref{fig:3} or the high-lifetime vesicles in Fig. \ref{fig:4} to be ignored in favour of global patterns – although the local prior will still retain these outliers. Consequently, the GP is most beneficial for samples that contain many examples of similar morphological structures, where these structures share similar lifetime properties. Since GPs capture sample dependent properties, they can be further exploited when the same sample is imaged across multiple regions of interest (for instance in a mosaic scan), by stacking intensity patches with corresponding lifetime labels into a common training set to improve generalisation.\\
The results demonstrate the fact that the introduction of the priors in the TV-based inverse retrieval algorithm, makes the latter a tractable problem. SiSIFUS gives the upsampled lifetime image sharp spatial features by extracting spatial information from an intensity image. This feature similarity is shown by perceptual metrics such as LPIPS, as features like edges, speckly textures and object shapes are captured by SiSIFUS, but not by interpolation. In contrast, pixelwise metrics penalise pixel-to-pixel estimation noise heavily, making them more lenient towards blurred, interpolated images than SiSIFUS. We therefore prefer the LPIPS metric (designed to measure image quality in a similar way to human perception) and used it to optimise our hyperparameters in validation. We note that Figures \ref{fig:4}, ref{fig:5}, and \ref{fig:6} were used as validation images that allowed us to optimise all hyperparameters (LP and GP window sizes, LP function, GP model architecture, epochs and learning rate, ADMM iterations and loss function weights). These were then fixed and used to generate the images in Fig. \ref{fig:3}.\\
It is worth emphasizing that SiSIFUS is currently not implemented simultaneously with high-speed measurements like the aforementioned 30 fps SPAD video, limiting its use for direct feedback or real-time diagnostics, such as intraoperative imaging. However, it enhances measurement speed by sampling fewer lifetime points and estimating lost information afterward. Equivalently, this mitigates phototoxicity and photodamage by illuminating the sample with less light, useful for imaging live samples in research or biopsy.\\
Despite these limitations, SiSIFUS offers notable potential for various applications beyond its current validation scope of FLIM. It could be applied to tasks involving disparate image types but exhibiting local or global correlations underscores its versatility and applicability in diverse research settings. \\
A key feature that we believe will be beneficial in any such approach is that image reconstruction is never based on statistical inference from other images - only the single image samples acquired from the two cameras are used thus strongly reducing or eliminating artefacts that may occur for example in other machine learned approaches that do indeed rely on large sets of additional data and images and thus representing a potential point of failure that is of concern in many applications.
\section{Materials and Methods}

\subsection{Experimental design}

\noindent{\bf{Mammalian cell culture conditions.}} Both the SKOV3 ovarian cancer cells and the Madin-Darby canine kidney (MDCK) cells were maintained in Dulbecco’s modified Eagle’s medium (DMEM) supplemented with 10\% FBS, 2 mM L-Glutamine and 1X PenStrep. Cell lines were maintained in 10 cm dishes at 37$^\circ$C and 5\% CO2.\\
SKOV3 cells were transfected in the morning using Amaxa Nucleofector (Lonza) kit V, program V-001 with either 5 $\mu g$ Raichu-Rac1-Clover-mCherry or pcDNA3.1-mClover DNA (adapted from \cite{itoh2002activation}) following manufacturers guidelines and replated on 6 cm TC-treated dishes at 37$^\circ$C and 5\% CO2. For live cell imaging, cells were collected and replated onto 35 mm glass bottom MatTek dishes that were previously coated overnight with laminin (10 $\mu g ml^{-1}$) diluted in PBS. These were left overnight at 37$^\circ$C, 5\% CO2.\\
The next morning prior to imaging, the dishes were washed twice with pre-warmed PBS and replaced with pre-warmed FluoroBrite DMEM supplemented with 10\% FBS, 2 mM L-Glutamine and 1X PenStrep. For fixed cell imaging, the cells were collected and replated onto 22 mm glass coverslips that were previously coated overnight with laminin (10 $\mu g ml^{-1}$) diluted in PBS. These were left overnight at 37$^\circ$C, 5\% CO2. The next day, these cells were fixed in 4\% PFA for 10 minutes and washed with PBS and mounted using Fluromount-G (Southern Biotech).\\
The MDCK cells were trypsinised and plated on 35 mm glass-bottom MatTek dishes and left to settle for 4 hours. Flipper-TR$^\circledR$ probe (Cytoskeleton; CY-SC020) was resuspended in 50 $\mu l$ anhydrous DMSO as per manufacturers instructions to yield a 1 mM stock. Flipper-TR was diluted in culture media to 2 µM and incubated on the cells overnight at 37$^\circ$C, 5\% CO2.\\
The next morning prior to imaging, the dishes were washed twice with pre-warmed PBS and replaced with prewarmed FluoroBrite DMEM (ThermoFisher Scientific; A1896701) supplemented with 10\% FBS, 2 mM L-Glutamine, 1X PenStrep and 2 µM Flipper-TR.\\
\noindent{\bf{Multiphoton raster-scanning time-domain FLIM: Experimental set-up details.}} For the dataset shown in Fig.5, cells were left to equilibrate on a heated microscope insert at 37$^\circ$C, perfused with 5\% CO2 prior to imaging. Images were acquired in the dark using a multiphoton LaVision TRIM scan head mounted on a Nikon Eclipse inverted microscope with a 20X water objective. Illumination is provided by a Ti:Sapphire femtosecond laser (Coherent Chameleon Ultra II) used at 920 nm (12\% power). The fluorescence signal was passed through band pass filters 525/50 nm emission and acquired using a FLIM X-16 Bioimaging Detector TCSPC FLIM system (LaVision BioTec). A 301 × 301 $\mu m^2$ FOV corresponding to 256 × 256 pixels was imaged at 600 Hz with a 10 line average in a total acquisition time of 5199 ms.\\
For the dataset shown in Fig. \ref{fig:3}, cells were left to equilibrate on a heated microscope insert at 37$^\circ$C, perfused with 5\% CO2 prior to imaging. Images were acquired in the dark using a multiphoton LaVision TRIM scan head mounted on a Nikon Eclipse inverted microscope with a Nikon Apo 60X oil objective, 1.4 NA. Illumination is provided by a Ti:Sapphire femtosecond laser used at 970 nm (8\% power) with an acquisition delay of 5.440 ns. The fluorescence signal was passed through emission band pass filters 600/60 nm and acquired using a FLIM X-16 Bioimaging Detector TCSPC FLIM system (LaVision BioTec).\\
A 163×163 $\mu m^2$ field of view correlating to 512×512 pixels was imaged at 600 Hz with a 70-line average for a total acquisition time of 72575 ms (High-Res). A total of 100 High and Low-Res images taken from 3 independent experiments. Background images (High and Low-Res) were obtained by closing the scan-head using the above settings. Instrument response function (IRF) was obtained using carbon nanorods with the above settings and a 1\% laser power.\\
\noindent{\bf{Widefield time-domain FLIM: Experimental set-up details.}} For the datasets shown in Figures \ref{fig:4} and \ref{fig:6}, a custom microscope system was built using high spatial resolution sCMOS sensor (Andor’s Zyla) and the FLIMera SPAD array sensor. Spatial registration was achieved by identifying a set of four co-registered points on the SPAD and CMOS, and mapping the CMOS image with a perspective transformation to match the field of view of the SPAD image. See the Supplement for a schematic of the experimental set up.
\subsection{Statistical Analysis}
\noindent{\bf{Inverse Retrieval Algorithm.}} The optimization is implemented using the alternating direction method of multipliers (ADMM) algorithm. For this the minimisation in Eq. \ref{eq:2} can be re-formulated as:
\begin{equation}\label{eq:5}
\begin{aligned}
\begin{split}
    \tau_{HR}{*} = & \underset{\hat{\tau}_{HR}}{\arg\min}\;C(\hat{\tau}_{HR}) \text{, where}\\
    C(\hat{\tau}_{HR}) &=\|\mathbf{A}\hat{\tau}_{HR}-\tau_{LR}\|_2^2+ \gamma\|\hat{\tau}_{HR}-\hat{\tau}_{LP}\|_2^2\\
    & + \beta\|\hat{\tau}_{HR}-\hat{\tau}_{GP}\|_2^2 +\alpha\|z\|_1\\
    & \text{subject to } \mathbf{A}\hat{\tau}_{HR}-z=0 \text{ and } \hat{\tau}_{HR}\geq 0
\end{split}
\end{aligned}
\end{equation}

The Augmented Lagrangian for this problem can be written as:
\begin{equation}\label{eq:6}
\begin{aligned}
\begin{split}
     L_\rho(\hat{\tau}_{HR}&,z,y) = \underset{\hat{\tau}_{HR}}{\arg\min}\;C(\hat{\tau}_{HR}) \text{, where}\\
    C(\hat{\tau}_{HR}) &=\|\mathbf{A}\hat{\tau}_{HR}-\tau_{LR}\|_2^2+ \gamma\|\hat{\tau}_{HR}-\hat{\tau}_{LP}\|_2^2\\
    & + \beta\|\hat{\tau}_{HR}-\hat{\tau}_{GP}\|_2^2 +\alpha\|z\|_1\\
    & + y\|\mathbf{D}\hat{\tau}_{HR}-z\| + \rho/2 \|\mathbf{D}\hat{\tau}_{HR}-z\|_2^2\\
\end{split}
\end{aligned}
\end{equation}
Here $y$ is the Lagrangian multiplier (or the dual variable) and $\rho$ is the penalty parameter. The ADMM approach involves jointly minimizing the Lagrangian over all the primal variables followed by the updates over the dual variables. The primal updates for the variables $\hat{\tau}_{HR}$ and z are given by:
\begin{align}\label{eq:78}
    \hspace*{-0.3cm} 
    \hat{\tau}_{HR_{k+1}}\leftarrow&\underset{\hat{\tau}_{HR}}{\arg \min}\|\mathbf{A}\hat{\tau}_{HR}-\tau_{LR}\|_2^2+ \gamma\|\hat{\tau}_{HR}-\hat{\tau}_{LP}\|_2^2\nonumber\\
    &+\beta\|\hat{\tau}_{HR}-\hat{\tau}_{GP}\|_2^2 + y_k(\mathbf{D}\hat{\tau}_{HR}-z_k) \nonumber \\
    &+\rho/2 \|\mathbf{D}\hat{\tau}_{HR}-z_k\|_2^2 \\
    z_{k+1}\leftarrow&\underset{z}\nonumber
    {\arg\min}\alpha\|z\|_1+y(\mathbf{D}\hat{\tau}_{HR}-z)\nonumber\\
    &+\rho/2\|\mathbf{D}\hat{\tau}_{HR_k}-z\|_2^2
\end{align}
The dual update is given by:
\begin{equation}\label{eq:9}
    y_{k+1}\leftarrow y_k + \rho(\mathbf{D}\hat{\tau}_{HR} - z)
\end{equation}
For the primal minimisation update, we use the standard optimization technique based on the fast iterative soft thresholding algorithm (FISTA). Each iteration of the ADMM hence comprises of 90 iterations of FISTA for the $\hat{\tau}_{HR}$ variable update.\\
The weighting factor $\gamma$ for the local prior term in the cost function has been kept constant for all the cases wherein $\gamma=1$. The factor $\beta$ on the other hand is varied for different upsampling factors such that it is 0.02 for 2x and 4x upsampling factors and 0.5 for higher upsampling factor of 8x and 16x. The GP cannot predict lifetimes within 6 pixels of the edges of the sample, since one cannot extract a 13x13 window centred on these pixels. Consequently, the GP’s contributions from these regions are removed from the IR reconstruction. A total of 20 iteration steps are used for minimizing the cost function - further iteration typically gives insignificant change in the solution. A Python implementation of the IR reconstruction for upsampling a 256 × 256 image to 512 × 512 (2x upsampling) takes approximately 80s.\\
\noindent{\bf{Local prior window size and function selection.}} In lieu of an analytical formula linking lifetime and intensity, the extent of the windows and the form of these local functions must be found empirically from the data. A set of window sizes, spanning from 2×2 to 8×8 were tested. Likewise, a set of different function forms were tried, including Gaussian Processes with RBF kernels, B-spline fitting, and Interpolation (Nearest, Linear, Cubic). The best form was found by comparing mean upsampling metrics over a validation set of four biological samples (shown in Fig. \ref{fig:3}, \ref{fig:4}, \ref{fig:5}, \ref{fig:6}) for four upsampling factors each (2x, 4x, 8x and 16x) - see Supplement for details. 5 × 5 windows and linear interpolation yielded the best results, hence these are used in all results shown in this work.\\
\noindent{\bf{Global prior data augmentation and training.}} Our global priors are generated from a neural network trained on a training set of intensity patches. The training set includes all patches with a central lifetime estimate, and their rotated and mirrored copies. Further, for high upsampling factors, their nearest neighbours are added to the training set (these neighbours are also in the test set), as well as the rotated and mirrored copies of these neighbours. Once trained, the network is used to evaluate all the original intensity patches, which includes all the training patches and all patches with unknown lifetimes. A 3×3 neighbourhood gives 9x more windows to train on; combined with augmentation for rotation and reflection invariance, this yields 72x more data. This dataset augmentation method assumes that within the pixel-pitch of the intensity image, lifetime varies slowly. This does not hold for every pixel, therefore, estimated labels have inherent uncertainty. Nonetheless, this form of blind labelling increases the diversity of the training input set. This was empirically found to be necessary for 8×8 and 16×16 upsampling not because of the upsampling factor, but rather because the decimated low-resolution input was so small, that the training set size was a severe limitation.\\
The 13x13 training and testing patches are copied into two channels, producing 13x13x2 input instances for the neural network. One channel is normalised on a per-patch basis, drawing focus to the shape and texture of the patch’s content. The other channel is divided by the maximum of the original intensity image, maintaining absolute intensity variations.\\
For the results shown in this paper, the network was trained 3 times with different random initialisations, and the prior of median quality was selected. We train on intensity patches 13×13×2 large, with ADAM \cite{kingma2014adam}, using a batch size of 100 over 150 epochs with mean absolute error (MAE) as the training loss, on an NVIDIA GeForce RTX 2080 Ti. Training on a 125 × 125 FLIM image (i.e. 125x125x8=125000 intensity patches due to 8× data augmentation for rotation and reflection invariance), takes $\sim$ 25 minutes of training on this hardware using TensorFlow, irrespective of the target image size. Trained DNNs tend to have negligible bias compared to the standard deviation of prediction. Since they are trained from scratch on each new sample, their training and validation losses vary from sample to sample, up-sampling factor to up-sampling factor, and initialisation to initialisation. Example test set performances can be seen in inset D of Fig. \ref{fig:3}, \ref{fig:4}, \ref{fig:5} and \ref{fig:6}.
\subsection*{Acknowledgements.}
\noindent{}The authors acknowledge funding from EPSRC (UK, grant no. EP/T002123/1), HORIBA Ltd., the Research Council of Lithuania (project No S-PD-24-35), the Royal Academy of Engineering (Chairs in Emerging Technology programme) and QuantIC (UK, grant no. EP/T00097X/1).\\
\subsection*{Data and materials availability.}
\noindent{}All data needed to evaluate the conclusions in the paper are present in the paper. Raw data, SISIFUS code, and analysis code can be found at \href{https://doi.org/10.5281/zenodo.10955467}{https://doi.org/10.5281/zenodo.10955467}.


%

\pagebreak
\onecolumngrid
\section*{Supplementary information}
\setcounter{section}{0}
\setcounter{figure}{0}
\renewcommand{\figurename}{Supplementary Fig.}
\section{Fluorescence intensity and lifetime}
\noindent{\bf{Lifetime and quantum yield.}} Fluorescence lifetime is described in literature as being independent of fluorescent intensity \cite{berezin2010fluorescence, szmacinski1995fluorescence}, and of fluorophore concentration \cite{suhling2015fluorescence} and excitation intensity. Here, we examine the context of these claims, and demonstrate the limitations of these generalisations.\\
In fluorescence, a photon excites a ground-state electron into an excited state, which then decays back to the ground state radiatively at a rate known as the decay rate. Other decay pathways compete with fluorescence, such as non-radiative decay and inter-system energy transfer between the fluorescent molecule and its environment. The probability of emitting a fluorescent photon per excitation event is called the quantum yield of fluorescence. Fluorescent intensity is the product of excitation intensity, the absorbance of the fluorophores (which depends strongly on their concentration) and fluorescence quantum yield.\\
The decay rate is the inverse of fluorescence lifetime, which is the expected time that an electron spends in the excited state before decaying via fluorescence. This is an intrinsic property of the molecule, and thus, is assumed to be independent of factors like fluorophore concentration. Consequently, fluorescence lifetime can be used to distinguish between different molecule populations.\\
However, fluorophores interact with their environment. The environment, in turn, can modulate both the excitation and emission pathways, changing both intensity and lifetime. Excitation can be enhanced or quenched by metallic surfaces or particles within the sample such as silver \cite{lakowicz2001radiative} via plasmonic resonance. Emission is modulated via nonradiative (or alternative) decay pathways, quenching the molecule’s radiative fluorescence as well as its lifetime, as derived in the main section of the paper.\\
\noindent{\bf{Fluorescence intensity.}} An imaging system generates a fluorescence intensity signal that depends on the spectral radiance $L_f(\lambda_o)$ of the sample and the net photon detection efficiency $PDE(\lambda_o)$ of the imaging system.\\
Let us consider a thin sample within the focal length of the optical system, using an epifluorescence setup. Using nomenclature from \cite{schwartz2002quantitating}, the spectral radiance $L_f(\lambda_o)[Wsr^{-1} m^{-2} nm^{-1})]$ emitted by the sample at wavelength $\lambda_o$ from excitation light at $\lambda_x$ is given by:
\begin{equation}\label{eq:s1}
    L_f (\lambda_o)=I_x N \Omega\epsilon(\lambda_x)Q(\tau,\lambda_o,\lambda_x)
\end{equation}
where $I_x$ is the incident excitation power $[W]$, $\tilde{N}(x,y)$ is the 2D concentration of fluorophores $[m^2]$ (the integral of the 3D concentration N of fluorophores along the length of the sample along the optical axis $z$, $\tilde{N}(x,y)=\int_z{N(x,y,z)dz}$, $\Omega$ is the solid angle through which emitted light is collected from the sample $[sr]$, $\epsilon(\lambda_x)$ is the absorptivity $[m^2]$ of the fluorophore as per the Beer-Lambert Law, and $Q(\tau,\lambda_o,\lambda_x)$ is the lifetime-dependent, spectral quantum yield of fluorescence.\\
This spectral radiance is imaged onto a detector that has a response $R[AW^{-1}]$ using a system with some étendue $\Gamma[m^2sr]$. To obtain the signal generated by the emission spectrum, we must integrate over the emission spectrum, giving:
\begin{equation}\label{eq:s2}
    s_o = \int_{\lambda_o}{L_f(\lambda_o)\Gamma R(\lambda_o)d\lambda_o}
\end{equation}
Substituting Eq. \ref{eq:s1} into Eq. \ref{eq:s2}, and integrating over the acquisition time $t_a$ gives us the measurement $M[C]$:
\begin{equation}\label{eq:s2}
    M = \int_{\lambda_o}\int_{t=0}^{t_a}{I_x\tilde{N}\Omega\epsilon(\lambda_x)Q(\tau,\lambda_o,\lambda_x)\Gamma R(\lambda_o)d\lambda_o dt}
\end{equation}
A fixed excitation and detection system allows us to calibrate the intensity $I_x$, the collection solid angle $\Omega$, the etendue $\Gamma$, the response $R(\lambda_o)$, and the acquisition time $t_a$. Therefore, variations of intensity across the field of view will depend on molecular concentration $\tilde{N}$ and absorptivity $\epsilon(\lambda_x)$ (whose product is the absorbance of the fluorophores), as well as the spectral quantum yield of fluorescence Q, which depends on fluorescence lifetime.\\
\noindent{\bf{Dependence of intensity on lifetime.}}
Absorbance and fluorescence lifetime appear to be unrelated, hence absorbance (ergo, fluorophore concentration) is an unpredictable confounding variable in intensity-lifetime dependencies.\\
Consequently, a fluorescence intensity measurement alone cannot give us full lifetime information. We therefore must use statistical priors to extract intensity-lifetime dependencies in the presence of biological confounding variables. A local prior is developed to extract dependencies when lifetime varies more rapidly in space than these confounding variables, or when they correlate with lifetime on local scales (either positively or inversely).\\
Further, many biological samples absorb fluorophores into particular subcellular compartments such as the cell membrane \cite{stockl2010detection}, vesicles \cite{pierzynska2014evaluation} or the nucleus \cite{estandarte2016use}. This results in lifetime patterns that often track cellular morphology. A global prior is developed to extract such dependencies. If absorbance were completely randomly distributed (which tends not to be the case in real samples), our method would not offer improvement over interpolation, instead our methods might overfit on noise patterns. To prevent this, our algorithm uses TV-filtering to prevent very unrealistically noisy lifetime estimates. \\
The question is whether recognisable intensity-lifetime dependencies actually exist in biological samples, or if absorbance renders them unusable. Below, we consider a series of case studies of fluorophore-environment interactions reported in literature, focusing on how these interactions modulate intensity and lifetime.\\
\noindent{\bf{Case studies.}}
Okabe et. al. \cite{okabe2012intracellular} used a complex fluorescent molecule made of a thermosensitive unit, a hydrophilic unit and a fluorescent unit to monitor temperature. In response to higher temperature, the molecule becomes hydrophobic, curling up and increasing both fluorescence quantum yield (thereby, intensity) and fluorescence lifetime. Fluorophore concentration still affects fluorescence intensity; however, locally (in regions of uniform concentration or at organelle edges), intensity and lifetime covary. Indeed, the authors use this probe to demonstrate temperature differences between the nucleus and cytoplasm of cells, which are visibly differentiable on both the lifetime and intensity maps.\\
Ogikubo et. al. \cite{ogikubo2011intracellular} used cellular auto-fluorescence of NADH to monitor intracellular pH. Their results show evident covariance of fluorescent intensity with fluorescence lifetime within cells; even though intensity is not a marker of pH, both intensity and fluorescence lifetime depend on the location of NADH within the cell. The reason for this is not explicitly explored, but different works have shown that the ratio of bound to free NADH depends on the local metabolism of the cell, which influences both the fluorescence lifetime and concentration of NADH autofluorescence \cite{datta2020fluorescence}. Correlations are similarly visible between NADH fluorescence intensity and lifetime in works by Stringari et. al. \cite{berezin2010fluorescence}, as both of these parameters are covariate with cellular redox ratio.\\
Van der Linden et. al. \cite{van2021turquoise} use FLIM as a tool for a quantitative measurement of calcium levels, independent of hardware. However, for a given hardware, fluorescent intensity spikes clearly show calcium spikes, even if they do not give absolute calcium concentrations on their own. Indeed, the authors demonstrate that their FLIM probe works by showing Supplementary videos of fluorescent intensity and lifetime side-by-side, which both show synchronised flickering. Lifetime and and intensity are strongly temporally correlated and are also locally correlated: cellular organoids have quasi uniform intensity and lifetime, both of which experience sudden gradients at organoid boundaries.\\
Verboogen et. al. \cite{verboogen2017fluorescence} demonstrate a FLIM-FRET probe for the imaging of SNARE trafficking in dendrites. For example, Förster resonance energy transfer (FRET) relies on this phenomenon. In FRET, the fluorophore, known as the donor, is linked to another fluorophore known as the acceptor, such that their relative conformation can change. The donor molecule is excited and its fluorescence measured. If the donor and acceptor are far, the donor will decay as if it were alone. If the donor and acceptor are in close vicinity, excited electrons can transfer energy from the donor onto the acceptor molecule, providing an alternative decay path for electrons, decreasing both fluorescence quantum yield (thus, intensity) and lifetime.\\
Gorpas et. al. \cite{gorpas2019autofluorescence} use skin autofluorescence to determine qualitative boundaries between cancerous and healthy skin tissue. They demonstrate that FLIM shows skin cancer; they do so by overlaying an augmented reality image of FLIM onto a visibly melanated patch of skin, whose colour correlates strongly with its lifetime.
\section{Local prior}
\noindent{}We performed a study to find the best window size and best function to map fluorescent intensity onto fluorescence lifetime with local priors. The window sizes were in the range 2 to 8, while the functions were a set of common schemes, ranging from B-splines (linear, quadratic and cubic), through regular interpolation (nearest, linear and cubic); and kriging (radial basis function Gaussian process fitting).\\
\begin{figure}[b]
    \centering
    \includegraphics[width=0.35\linewidth]{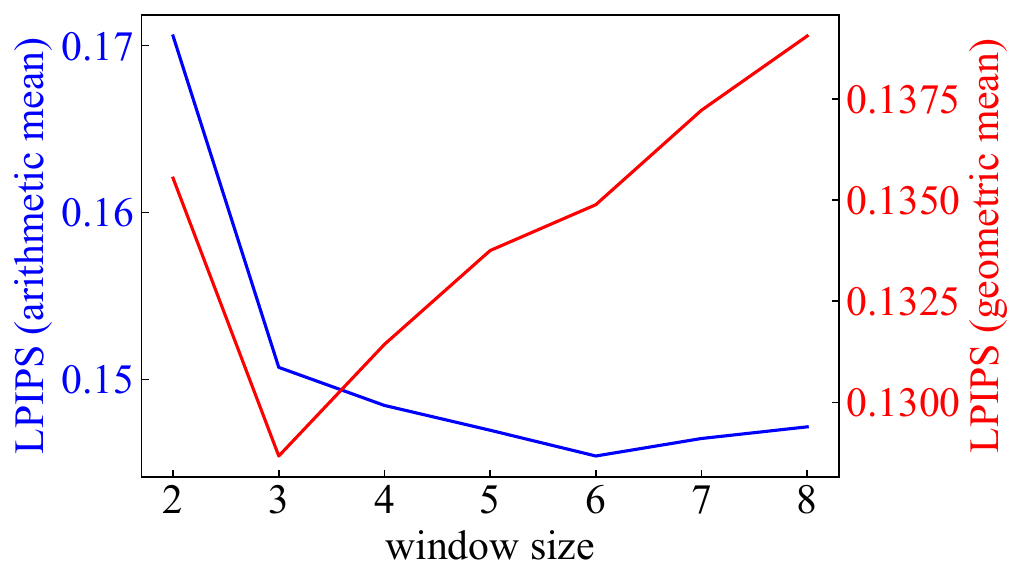}
    \caption{We found the mean LPIPS for priors generated using various window sizes, averaged across our 4 samples and 4 upsampling factors (2,4,8,16). We plot both the geometric and arithmetic mean.}
    \label{fig:s2}
\end{figure}
\begin{figure}[t]
    \centering
    \includegraphics[width=0.6\linewidth]{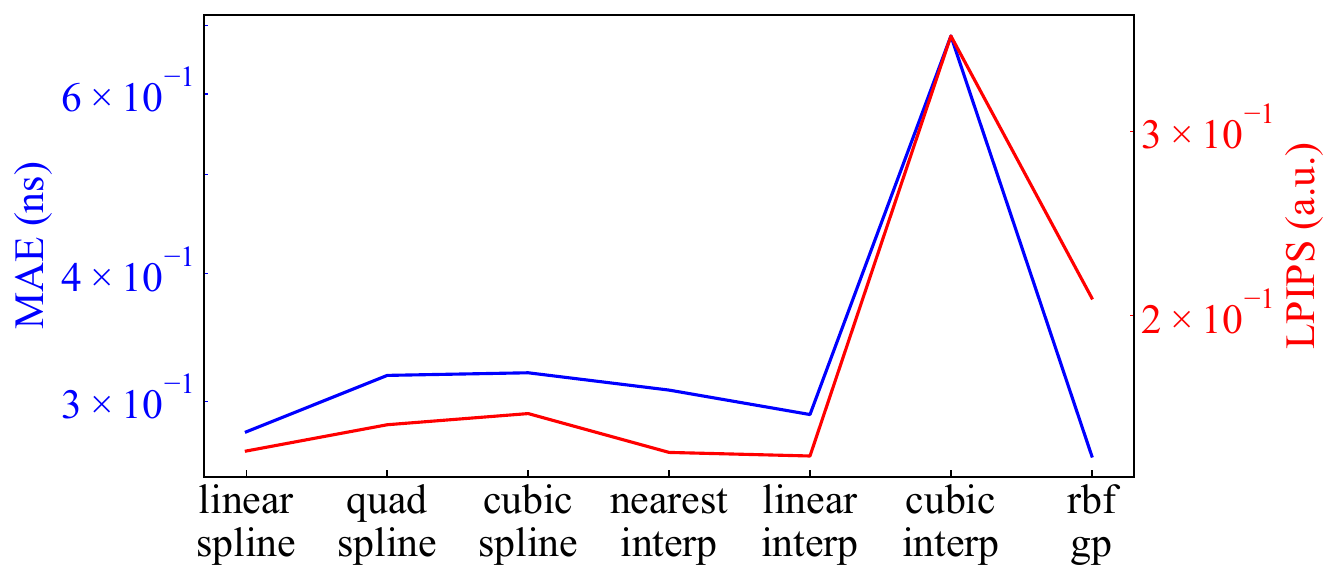}
    \caption{We found the mean MAE and LPIPS for priors generated using various LCP types, averaged across our 4 samples and 4 upsampling factors (2,4,8,16).}
    \label{fig:s3}
\end{figure}\\
We applied these window sizes and functions to 4 samples (including the three shown in Figures \ref{fig:4}, \ref{fig:5}, and \ref{fig:6}) and 4 upsampling factors (2,4,8, and 16x). We evaluated the methods based on mean-absolute-error and LPIPS between the reconstruction and ground truth, averaged over these 4×4 scenarios. Our results are shown in Supplementary Fig. \ref{fig:s2} and Supplementary Fig. \ref{fig:s3}. Based on these results, we decided to use a window size of 5 and linear interpolation for generating LPs.
\begin{figure}[b]
    \centering
    \includegraphics[width=0.75\linewidth]{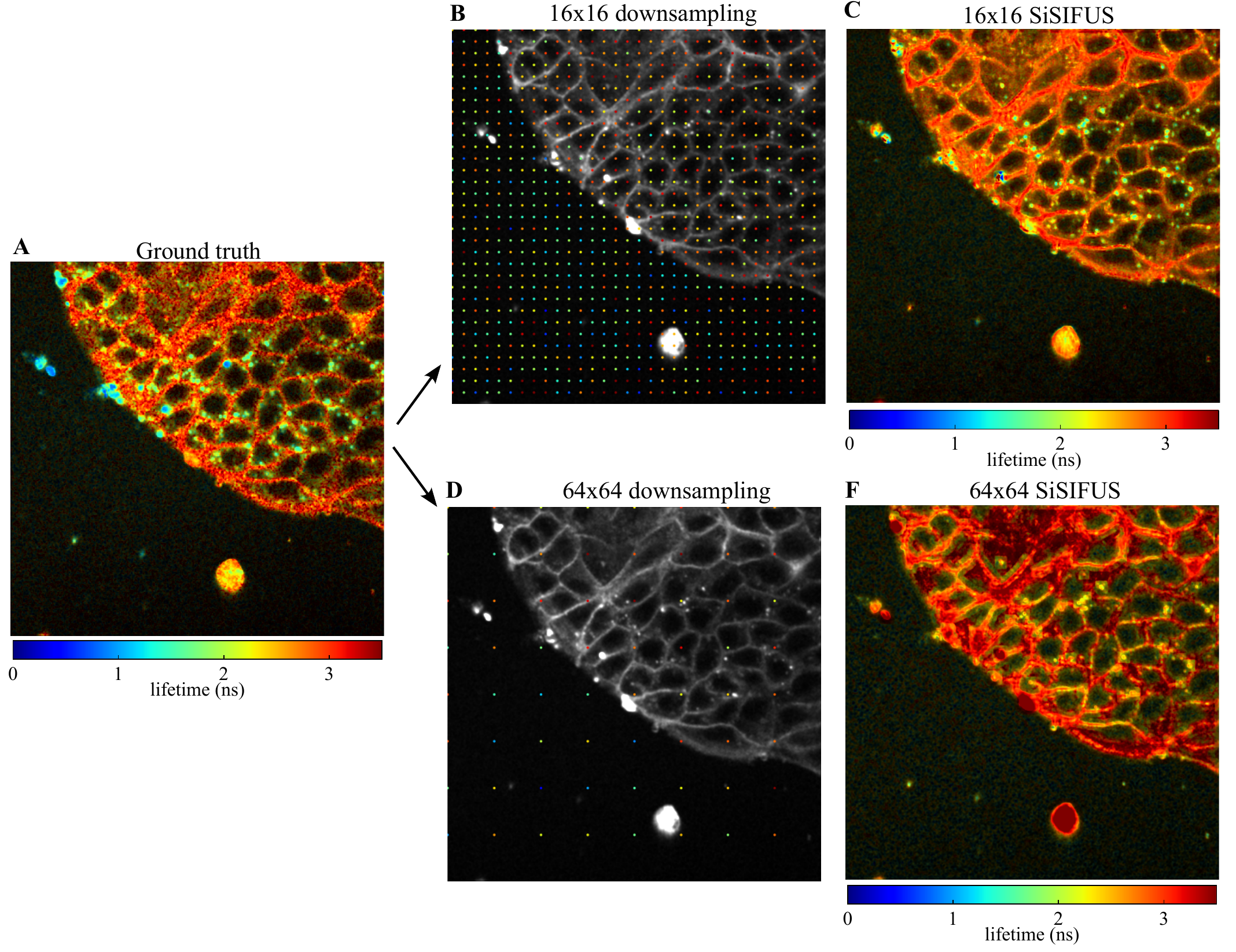}
    \caption{\textbf{Illustration of the impact of under-sampling on image resolution.} \textbf{(A)} The original image is 512x512 with a pixel pitch of 0.33$\mu m$. \textbf{(B)} Data is decimated by 16x16, with lifetime samples overlaid. \textbf{(C)} SiSIFUS can reasonably reconstruct the sample from intensity-lifetime pairs. \textbf{(D)} The ground truth lifetime is instead decimated by 64x64, resulting in only 8x8 lifetime measurements. \textbf{(E)} This extreme under-sampling causes SiSIFUS to fail in accurately recovering the lifetime distribution.}
    \label{fig:s1}
\end{figure}\\
\section{Sampling considerations}
\noindent{}SiSIFUS relies on either local dependencies, where there are gradual changes in confounding variables such as fluorophore concentration, or global dependencies between structure and lifetime. Local dependencies involve scenarios like free fluorophores in the cytoplasm, which diffuse to achieve a locally uniform density, or beads coated in fluorescent dyes. Another example is continuous cell membranes treated with diffusive dyes like Flipper TR.\\
Global dependencies assume that objects with similar shapes in the FOV have similar lifetimes, as illustrated in Fig. \ref{fig:3} and Fig. \ref{fig:5}. For instance, in a scenario with a mixture of small and large fluorescent beads coated with different fluorophores, by sparsely sampling their lifetimes, we can infer the lifetimes of other beads in the image. Employing machine learning algorithms, SiSIFUS automates and enhances this pattern-matching process, foregoing manual matching.\\
Capturing these dependencies is crucial for SiSIFUS; thus, sampling density must be adequate to measure them. Whether sampling locally or across the FOV, denser sampling is required for small or rare biological or mechanical structures. In general, the sparser the structures in the FOV and the greater the variety of distinct lifetimes expected to be resolved, the denser the sampling required. In the bead example, if a medium-sized bead is present but never sampled for lifetime, accurate estimation requires external knowledge. We illustrate the effect of undersampling with a 64x64 super-resolved example of the MDCK-Flipper TR sample in Fig. \ref{fig:3} of the main text (from 8x8 to 512x512). Results are shown in Supplementary Fig. \ref{fig:s1}.
\section{Visualisation}
As stated in the main paper, we visualise our FLIM data by overlaying (weighting) it with local contrast enhanced intensity. This allows us to see lifetime patterns more clearly than using the raw lifetime image. For this, we first choose a colormap to plot the lifetime data in and use this colormap to convert it into an RGB image.\\
Separately, we apply Contrast Limited Adaptive Histogram Equalization (CLAHE) to the intensity image. This makes it to that the image has good contrast across the field of view, without saturating bright spots or rendering dark regions imperceptible. We then scale each channel of the RGB image with this contrast enhanced intensity, preserving the RGB image’s color, but manipulating its brightness.
\section{Metric details}
Pixelwise metrics such as mean absolute error (MAE), mean squared error (MSE) and peak signal-to-noise ratio (PSNR) are commonly used in image processing. Compared to MAE and MSE, PSNR is adjusted for image scale, letting it generalise image similarity across lifetime maps of different ranges. PSNR is guaranteed to favour the same method as MSE for a given sample, whilst allowing us to compare different samples as well. A quantitative and straight-forward pixel-to-pixel comparison, PSNR is valuable for automated tasks, however, a widely known issue of  metrics is how poorly they reflect the human perspective of image similarity. A particularly infamous example is blurring, which produces relatively low pixelwise errors even when deteriorating image quality severely.\\
\begin{figure}[b]
    \centering
    \includegraphics[width=0.9\linewidth]{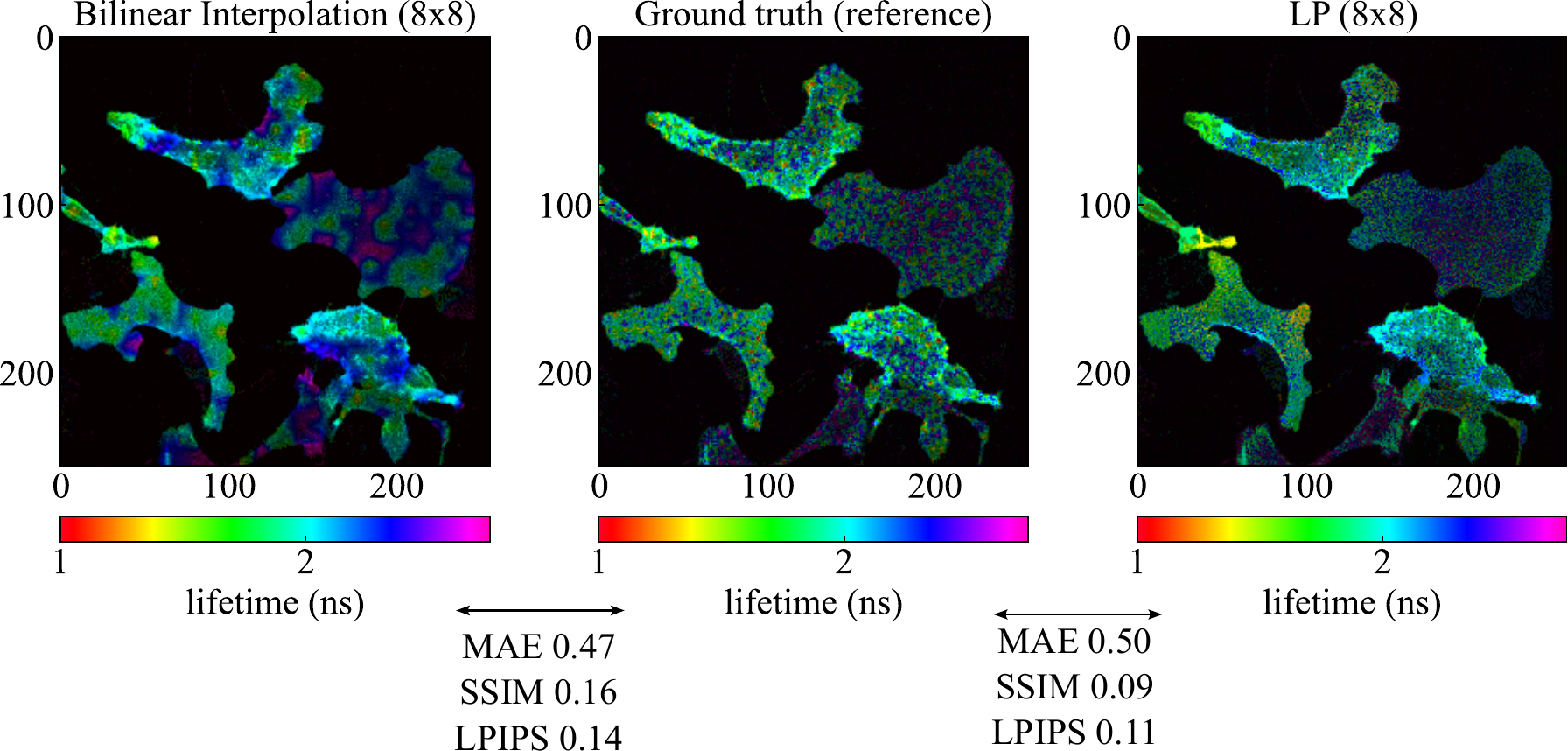}
    \caption{We show a comparison of interpolation vs an LP for 8 × 8 upsampling of a Rac-Raichu sample. Left to right: bilinear upsampling; the ground truth lifetime; LP. Both MAE and SSIM favour interpolation over the LP. However, the interpolated image is blurred and has artefacts matching the low-resolution sample grid, so, a user would likely say the LP is closer to the ground truth. LPIPS captures this perceptual similarity, favouring the LP.}
    \label{fig:s4}
\end{figure}\\
To address this, simple perceptual metrics such as the structural similarity index measure (SSIM) and multiscale SSIM (MS-SSIM) have been proposed. Indeed, SSIM is more sensitive to blurring than MSE and less sensitive to noise \cite{dosselmann2011comprehensive}, akin to human perception. That said, SSIM and MS-SSIM are strongly correlated with MSE; in fact, their ‘luminance’ term is equivalent to MSE. Further, a statistical evaluation \cite{sheikh2006statistical} found that, while SSIM correlates with human perceptual score to a factor of 0.9393, the PSNR (equivalently, MSE) baseline yields a score of 0.8709. This suggests that structural metrics, while an improvement on pixelwise metrics, still can be improved upon. In our study, we evaluate SSIM on windows of size 25.
One of the early realisations regarding convolutional neural networks was that the first few convolutional layers in a machine learning architecture tend to perform feature extraction, of growing abstractive complexity with increasing layer depth. This understanding led to the development of learned perceptual image patch similarity (LPIPS) by Zhang et. al. \cite{zhang2018unreasonable}. The authors discovered that distance in the feature space of some image-trained neural network correlates well with human perception. Since then, LPIPS has gained widespread traction, for instance the recently published image-to-image and text-to-image benchmark, Stable Diffusion, uses LPIPS in training its autoencoder structure \cite{rombach2022high}. The authors have added a Python implementation via the lpips package \cite{lpips} including 3 pre-trained neural networks (In our works, we use the AlexNet network). LPIPS is open-source and provides fixed-weight neural networks with a fixed protocol for calculating image similarity, which makes our LPIPS values exactly replicable. \\
We show an example from experimental data to illustrate the disparity between human vision and metrics with pixelwise components in Supplementary Fig. \ref{fig:s4}.
\section{Experimental setup details}
The SPAD array datasets shown in Fig. \ref{fig:2} and Fig. \ref{fig:5} of the main paper were obtained by our bespoke microscope system comprising of the 192×128 pixels SPAD array sensor (FLIMera) and the sCMOS sensor (Andor Zyla), shown in Supplementary Fig. \ref{fig:s5}. The system is a widefield epifluorescence setup, observing the sample using a 60x 1.4NA Nikon oil objective, alongside a 250mm focal length tube lens for the Zyla sensor and a 89.9mm focal length tube lens for the SPAD sensor. In this arrangement, we obtain 192 × 128 × 326 datacube for the dataset shown in Fig. \ref{fig:2}. However, the detector’s pixel layout is such that two-columns of SPAD active areas are followed by a ‘dead space’ which is two pixel-pitches wide. For the dataset shown in Fig. \ref{fig:5} of the main paper, we correct for this irregular sampling. The full FOV is sampled in two shots, such that the image plane is translated by two pixel-pitches between the shots, and the outputs are then fused computationally. This yields a fluorescence datacube of size 192 × 256 × 326.
\begin{figure}[H]
    \centering
    \includegraphics[width=0.7\linewidth]{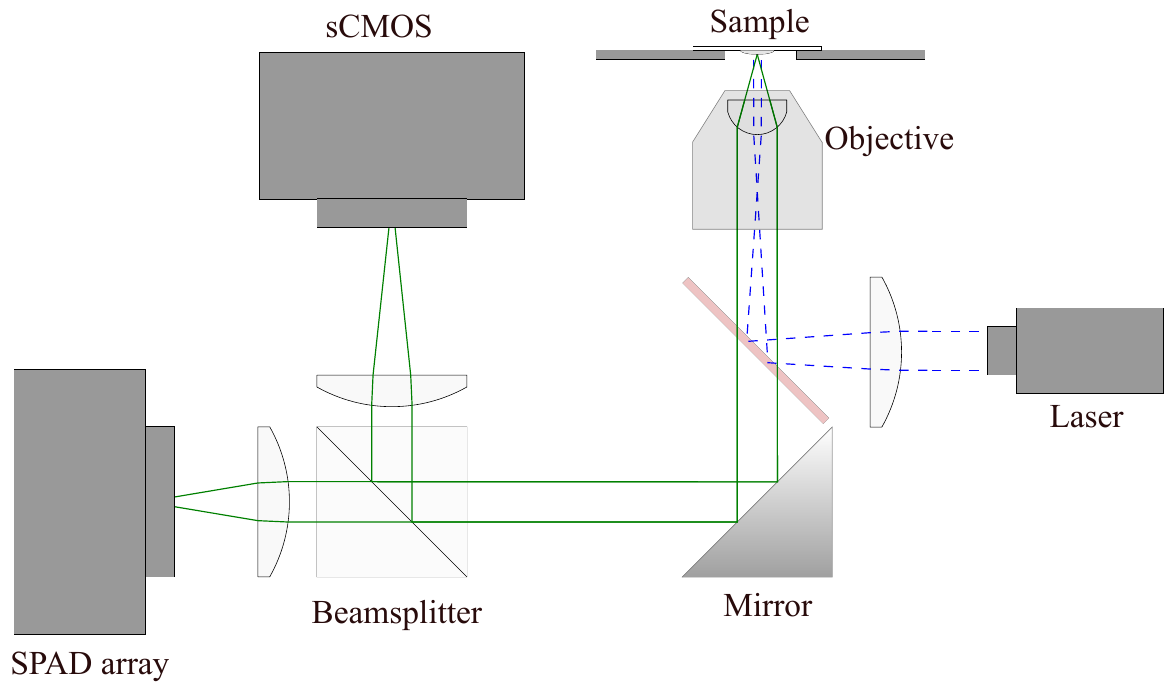}
    \caption{Schematic representation of the experimental set up: the sample is illuminated by a pulsed laser. The fluorescence signal from the sample is split by a dichroic beamsplitter, then collected and imaged on to the high spatial resolution sCMOS sensor and the low spatial resolution SPAD array sensor.}
    \label{fig:s5}
\end{figure}


\end{document}